\documentclass[
  ,draft            % use draft while you are working on the paper
]
  {aipproc}

\layoutstyle{6x9}

\newcommand{\tr}{{\textrm {tr}}}

\newcommand{\D}{{\widehat D}}

\newcommand{\cL}{{\cal L}}

\def\slashchar#1{\setbox0=\hbox{$#1$}
   \dimen0=\wd0 \setbox1=\hbox{/} \dimen1=\wd1
   \ifdim\dimen0>\dimen1 \rlap{\hbox to \dimen0{\hfil/\hfil}} #1
   \else  \rlap{\hbox to \dimen1{\hfil$#1$\hfil}} / \fi}

\def\D{{\bf D}}

\begin{document}

\title{Chiral Lagrangians at finite temperature and the Polyakov Loop
\thanks{Presented by E.R.A. at the Eurocongress 
on ``Quark Confinement 2004'', Cagliari, Sardinia, 21-25 September-2004.}}

\author{E. Meg\'{\i}as}{
address={Departamento de F{\'{\i}}sica Moderna,  
Universidad de Granada. 18071-Granada (Spain) }
}
\author{\underline{E.Ruiz Arriola}}
{
address={Departamento de F{\'{\i}}sica Moderna,  
Universidad de Granada. 18071-Granada (Spain) }
}
\author{L.L. Salcedo}{
address={Departamento de F{\'{\i}}sica Moderna,  
Universidad de Granada. 18071-Granada (Spain) }
}

\begin{abstract}
Heat kernel expansions at finite temperature of massless QCD and chiral
quark models generate effective actions relevant for both low and high
temperature QCD. The key relevance of the Polyakov Loop to maintain the
large and non-perturbative gauge invariance at finite temperature is
stressed.
\end{abstract}

\maketitle

%%%%%%%%%%%%%%%%%%%%%%%%%%%%%%%%%%%%%%%%%%%%
%% MAINMATTER
%%%%%%%%%%%%%%%%%%%%%%%%%%%%%%%%%%%%%%%%%%%%

%\subsection{Compositness and Finite Temperature}
%\label{sec:intro}
In the imaginary time formulation of quantum field
theory~\cite{Landsman:1986uw}, finite temperature is introduced by
imposing periodic or anti-periodic boundary conditions for bosons and
fermions respectively. This approach necessarily breaks the Lorentz
invariance of the starting Lagrangian since the heat bath is a static
infinitely heavy system which is assumed to be at rest and communicates
energy at no expense with the fundamental degrees of freedom. This
fact generates an unpleasant plethora of possible Lorentz
contributions to Feynman diagrams which fortunately are ultraviolet
finite due to the presence of suppressing kinetic Boltzmann factors. As
a gratifying consequence, the finite temperature renormalizability of
a theory relies indeed on its renormalizability at zero temperature
and provides any fundamental theory with definite predictive power, if
the parameters entering the Lagrangian are assumed to be temperature
independent. This standard procedure might be called {\it minimal
thermal coupling} of the heat bath since, at least in perturbation
theory, finite temperature (and hence Lorentz breaking) effects are
indeed quantum effects. For an effective field theory of composite
particles the situation may be not that simple since the corresponding
Low Energy Constants (LEC's) encode the microscopic information of the
underlying substructure. Actually, the constituents are also heated up
and it is clear that LEC's inherit a finite temperature
dependence. Moreover, genuinely Lorentz breaking and temperature
dependent terms might be further added to the effective
Lagrangian. 

The previous discussion is actually relevant to QCD at finite
temperature~\cite{Gross:1980br,Svetitsky:1985ye}.  At low energies and
temperatures, one uses Chiral Perturbation Theory
(ChPT)~\cite{Gasser:1984gg} and assumes that in the effective theory
the LEC's are temperature independent; the $T$ dependence is generated
through thermal pion loops~\cite{Gasser:1986vb}. This assumption
relies strongly on the existence of a mass gap in the physical
hadronic spectrum due to the presence of would-be massless
pseudo-scalar Goldstone bosons, so one expects the leading temperature
effects, well below the phase transition, to be ${\cal O} ( e^{-M_\pi
/T} )$, whereas the next corrections might be ${\cal O} ( e^{-M_\rho
/T} )$. At present, one cannot compute the LEC's from QCD itself, but
some insight on the base of chiral quark models can be
gained. Actually, the way how these Boltzmann factors 
actually arise in chiral quark models is not at all trivial.

To illustrate the occurrence of non-minimal thermal couplings in
effective Lagrangeans, we have used based on previos
work~\cite{Salcedo:1998sv,Garcia-Recio:2000gt} the heat kernel
method~\cite{Megias:2002vr} (see also
Ref.~\cite{Megias:2003ui,Megias:2004bj}) for an application to high
temperature QCD) and computed
recently~\cite{Megias:2004kc,Megias:2004} the chiral Lagrangian using
chiral quark models where pseudo-scalar Goldstone bosons arise as
bound $q\bar q $ states. At finite temperature it can be written in
the standard Gasser and Leutwyler~\cite{Gasser:1984gg}
form~\footnote{We use an asterisk as upper-script for finite
temperature quantities, i. e. ${\cal O}^* = {\cal O}_T $.}
\begin{eqnarray}
%\cL^*_q{}^{(0)} &=& -\frac{2N_f}{(4\pi)^2}
%\langle\tr_c\J_{-2,0}(\Lambda,M,\hat\nu)\rangle \,, \\
\cL^*_q{}^{(2)} &=& \frac{{f^*}^2}{4}\tr_f\left( \D_\mu
U^\dagger\D_\mu U +(\overline\chi^\dagger U +\overline\chi U^\dagger)
\right) \,, \nonumber \\ \cL^*_q{}^{(4)} &=& \sum_{i=1}^{10} L_i^*
{\cal L}_i +\sum_i L_i' {\cal L}'_i 
\label{eq:lag_4}
\end{eqnarray}
Here, $f^*$ is the pion weak decay constant at finite $T$  in
the chiral limit, and is given by
\begin{eqnarray}
{f^*}^2 &=& 4 M^2 \, T \, {\rm Tr}_c \sum_{\omega_n}
\int \frac{d^3 k}{(2\pi)^3} \frac1{\left[\omega_n^2+ k^2 + M^2
\right]^2}.
\label{eq:fpi} 
\end{eqnarray}
with $M$ the constituent quark mass and $\omega_n= 2 \pi T ( n+1/2 ) $
the fermionic Matsubara frequency and ${\rm Tr}_c $ is the colour
trace. The constants $L_{1,\dots, 10}^*$ appearing in the higher order
terms are displayed in Ref.~\cite{Megias:2004} are the temperature
dependent LEC's of the Lorentz preserving terms whereas there are
appear new ones which break explicitly the symmetry and obviously
vanish at zero temperature. Note that at lowest order in the chiral
counting there is no Lorentz breaking terms.  These expressions
illustrate our point, that the LEC's may depend themselves on the
temperature. On the other hand, it also raises a puzzle, because the
leading thermal effects in the large $N_c$ limit are ${\cal O} ( N_c
e^{-M/T} )$ (see also Ref.~\cite{Florkowski:1996wf}), but in ChPT
there are no leading large $N_c$ corrections.

The solution to the puzzle has to do with the fact that for gauge
theories like QCD at finite
temperatures~\cite{Landsman:1986uw, Gross:1980br, Svetitsky:1985ye} the
non Abelian gauge manifest non perturbatively.  In the Polyakov gauge,
where $ \partial_4 A_4=0$ and $ A_4 $ is a diagonal and traceless $N_c
\times N_c $ matrix, and $N_c$ is the number of colors, there is still
some freedom in choosing the gluon field. Under periodic gauge
transformations~\cite{Salcedo:1998sv,Garcia-Recio:2000gt} the
requirement of gauge invariance really implies identifying all Gribov
replicas  differing by a multiple of $2 \pi / \beta $, which means
periodicity in the diagonal amplitudes of $A_4$ of period $ 2 \pi /
\beta$. Perturbation theory, which corresponds to expanding in powers
of small $A_4$ fields manifestly breaks gauge invariance at finite
temperature.  A way of avoiding this explicit breaking is by
considering the Polyakov loop $\Omega$ as an independent variable,
\begin{eqnarray}
\Omega  = e^{{\rm i} \beta A_4 (\vec x)}
\end{eqnarray} 
We have recently developed an expansion keeping these symmetries in
general theories and applied it to QCD at the one quark+gluon loop
level~\cite{Megias:2002vr,Megias:2003ui}. When these ideas are
incorporated into the chiral quark model framework at the leading one
quark loop approximation there appears an accidental center symmetry
similar to the one of QCD in the quenched approximation, under
aperiodic gauge transformations~\cite{Svetitsky:1985ye} of the center
$Z(N_c)$ of the group $SU(N_c) $.  The Polyakov loop transforms as the
fundamental representation of the $Z(N_c)$ group, and hence $\langle
\Omega \rangle =0 $ in the unbroken center symmetric and confining
phase.  More generally, $ \langle \Omega^n \rangle =0 $ for $ n \neq m
N_c $ with $m$ an arbitrary integer. The anti-periodic quark fields at
the end of the Euclidean imaginary interval transforms also as the
fundamental representation of the center group, so that the center
symmetry is explicitly broken by the presence of dynamical quarks, and
hence in the quenched approximation non-local condensates fulfill a
selection rule of the form, $ \langle \bar q ( n \beta ) q( 0) \rangle
=0 $ for $ n \neq m N_c $. This selection rule has some impact on
chiral quark models. In practice, the Polyakov loop can be coupled to
the chiral quark model using the modified fermionic Matsubara
frequencies \cite{Salcedo:1998sv,Garcia-Recio:2000gt}
\begin{eqnarray}
\hat \omega_n = 2 \pi T ( n+1/2 + \nu)\,, \quad \nu=(2\pi i)^{-1}\log\Omega
\end{eqnarray}
which are shifted by the logarithm of the Polyakov loop which we
assume for simplicity to be $\vec x$ independent, as suggested also in
Refs.~\cite{Gocksch:1984yk,Meisinger:2002kg,Fukushima:2003fw}. The
${\rm Tr}_c $ in Eq.~(\ref{eq:fpi}) does not simply give $N_c$ because
this coupling introduces a colour source into the problem for a fixed
$A_4$ field. After projection onto the colour neutral states by gauge
averaging, at the one quark loop level, there is an accidental
$Z(N_c)$ symmetry in the model which generates a similar selection
rule as in pure gluodynamics, from which a strong thermal suppression,
$ {\cal O} (e^{-N_c M /T } )$ follows in agreement with a previous
observation~\cite{Oleszczuk:1992yg}.  In this way compliance with ChPT
at finite temperature can be achieved.

%\smallskip 
%\begin{acknowledgments}
{\sl This work is supported in part by funds provided by the Spanish DGI
with grant no. BMF2002-03218, Junta de Andaluc\'{\i}a grant no. FM-225
and EURIDICE grant number HPRN-CT-2003-00311.} 
%\end{acknowledgments}

\end{document}